\documentclass[prd,preprint,aps,dvips,showpacs]{revtex4}
\usepackage{graphicx}
\usepackage[english]{babel}
\usepackage{amscd}
\usepackage{epsfig}
\usepackage{tabularx}
\usepackage{latexsym}
\usepackage{amsmath}
\usepackage{amsfonts}
\usepackage{amssymb}
\usepackage[latin1]{inputenc}
\usepackage{epsfig}
\usepackage{times}
\usepackage[T1]{fontenc}
\usepackage{graphics}
\usepackage{verbatim}
\usepackage[absolute]{textpos}
\usepackage{wrapfig,times}
\usepackage{amsthm}
\usepackage{setspace}

\newcommand{\be}{\begin{equation}}
\newcommand{\ee}{ \end{equation}}
\newcommand{\ben}{\begin{eqnarray}}
\newcommand{\een}{\end{eqnarray}}

\begin{document}

\title{Hybrid deformed algebra}

\author{Andre A. Marinho$^{1}$, Francisco A. Brito$^{1,2}$}

\affiliation{$^{1}$ Departamento de Física, Universidade Federal de
Campina Grande, 58109-970 Campina Grande, Paraiba, Brazil
\\
$^{2}$ Departamento de Física, Universidade Federal da Paraíba,
Caixa Postal 5008, 58051-970 João Pessoa, Paraíba, Brazil}
\date{\today}

%%%%%%%%%%%%%%%%%%%%%%%%%%%%%%%%%%%%%%%%%%%%%%%%%%%%%%%%%%%%%%%%%%%%%%%%%%%%%
\begin{abstract} 

By considering $p,q$-deformed  and $\mu$-deformed algebras we propose an association of them to form a hybrid deformed algebra. The increased number of 
available parameters can provide us with a richer tool to investigate new scenarios within hybrid deformed statistics.

\end{abstract}


\maketitle

%%%%%%%%%%%%%%%%%%%%%%%%%%%%%%%%%%%%%%%%%%%%%%%%%%%%%%%%%%%%%%%%%%
\section{Introduction}

In conducting a study on basic hypergeometric series at the beginning of the last century, the mathematician F.H. Jackson 
developed what we call $q$-calculus and its famous Jackson derivatives (JD) \cite{jak}. The emergence of deformed algebra 
played a central role in the representation of quantum groups \cite{bie,mac,fuc,erz}. A concrete realization very widespread in 
the literature is the deformation of the Heisenberg algebra with the introduction of the parameter of deformation $q$ through  
$q$-oscillators \cite{arik,jan,ext,kul,gas,bur,dam,tru,erns,bon,anc,lav,kac,shu,ere,aba1,amg2,mir,zar,bri,dey,fak,nut}, which 
obey the commutation relations that define the Lie algebra, and that in the limit $q\to 1$ we recover the original algebra.  This new 
mathematical structure immediately attracted interest due to its relevance to various problems in theoretical physics which includes, e.g., 
cosmology and condensed matter physics. 

The statistical mechanics of bosons and fermions are strongly linked to microscopic thermodynamics, that in macroscopic level we can describe these 
systems through macroscopic variables such as pressure and internal energy. The generalization comes naturally through 
intermediate statistics \cite{aro,wil,che,ach,perk,nar2,lav1,rov,aba3} and non-standard quantum statistics
\cite{gen,gre,pol,abe} as well as non-extensive models \cite{tsa,kan,arp,ben,bag,gom,abr}. 
In addition to the $q$-deformation parameter, the insertion of two parameters $p,q$ \cite{chak,bur1,aba4,jin,sou,amg1,chu1,sch,buk,per} 
the $q_1$ and $q_2$-deformed Fibonacci oscillators \cite{arik1,aba,bri1,bri2}, as well as multiple parameters 
\cite{chu,arik2,aba2,ssm,bor,bur2,hou,amg3,su} were proposed in the literature. 

We also emphasize another type of deformation which, as we shall see, presents the same structure of the $q$-algebra, but with the 
insertion of a new $\mu$ parameter that leads to the $\mu$-deformation \cite{jan1,amg} which presents a new definition of the so-called 
\textit{basic number}. As we shall show, all these proposals allow us to generalize the polynomials and the 
Fibonacci sequence. Quantum algebra with two (or more) deformation parameters may have greater flexibility when dealing with 
realistic (phenomenological) applications in physical models.

In this paper, we describe the deformed oscillator models intensively investigated in literature by Jackson, Biedenharn-Macfarlane, 
Arik-Coon and $p,q$-oscillators \cite{jak,bie,mac,chak,arik1}, to show that all deformed algebras satisfy Fibonacci relations. For this, we will start 
with the generalized definition of the \textit{basic number} for the Fibonacci oscillators, such that when we take the appropriate values for 
$q_1$ and $q_2$ we describe the aforementioned models. 

For the sake of generality, we are led to propose a new model, where we associate 
the well-known $q$-deformed models with the $\mu$-deformation \cite{jan1,amg}. This will produces a {\it hybrid model} that allows more flexibility 
in handling it, that is, we can insert at the same time one, two or three deformation parameters to a system, 
and by taking the proper limits the model returns to standard ones. We can apply the hybrid model
in various known areas of physics, such as in recent works using standard models of deformation, e.g., in the 
 black holes  physics \cite{tsa1}, dark energy and dark matter \cite{eba,amg4,dil,tav}.

The paper is organized as follows. In Section \ref{dha} we generalize the Heisenberg algebra by initially using the Fibonacci 
oscillators and in the sequel obtaining the other models of ($q$ or $p,q$)-deformation. In Section \ref{mdo} 
we define our hybrid deformation model and finally, in Section \ref{con} we make our final comments.

%%%%%%%%%%%%%%%%%%%%%%%%%%%%%%%%%%%%%%%%%%%%%%%%%%%%%%%%%%%%%%%%%%%%%%%%%%%%%%%%%%%%%%%%
\section{Deformed Heisenberg Algebra}
\label{dha}

We will start with using the so-called Fibonacci oscillators, however, it is worth noting that the oscillators that will be described 
below are all accommodated within the same mathematical structure of the deformed generalized oscillator, which is defined 
by Heisenberg algebra in terms of operators of annihilation and creation ($c$ and $c^\dagger$, respectively), 
of the operator number $N$ and structure function $\Phi(x)$, satisfying the relations
\be \label{e1} c_{i} c_{i}^\dagger - q_1^{2}c_{i}^\dagger c_i = q_2^{2n_i}\qquad\mbox{or}\qquad c_i c_{i}^\dagger - 
q_2^{2}c_{i}^\dagger c_i = q_1^{2n_i},\ee
\begin{equation}\label{e1.1}[N,c^{\dagger}] = c^{\dagger}, \qquad\qquad [N,c] = -c. \end {equation}
\be \label{e2} c_{i}^\dagger c_{i} = \Phi(N) = [N],\qquad\qquad c_{i}c_{i}^\dagger = \Phi(N+1) = [N+1],\ee
where $\Phi(x)$ is a characteristic positive analytical function for the deformation regime, with $\Phi(0) = 0$.
We conclude from Eq.~(\ref{e2}) that for $N=\Phi^{-1} (c_{i}^\dagger c_{i})$, the commutation and anti-commutation relations are satisfied, i.e., 
\be [c_{i},c_{i}^\dagger] = [N+1] - [N],\qquad \{c_{i},c_{i}^\dagger\} = [N+1] + [N].\ee

The commutation relations (\ref{e1}) are valid for the definition of the Fibonacci \textit{basic number or $n$-bracket} \cite{arik1}
\be \label{e3}[n_{i,q_1,q_2}] = c_{i}^\dagger c_{i} = \frac{q_1^{2n_i}-q_2^{2n_i}}{q_1^2-q_2^{2}},\ee
where $q_1$ and $q_2$ are parameters of deformation that are real, positive, independent and at the 
limit $q_1=1$ and $q_2\to 1$ (or vice versa) we have to $[n_{i,q_1,q_2}]=n_i$. 

The Fock space spanned by the orthonormalized eigenstates $|n \rangle$ is constructed according to
\begin{eqnarray} {|n\rangle} =\frac{(c^{\dagger})^{n}} {\sqrt{[n]!}}{|0\rangle},\qquad\qquad c{|0\rangle}=0 ,\end {eqnarray}
and the actions of $c$, $c^{\dagger}$ and $N$ on the states $|n\rangle$ in the Fock space are known to be
\begin{equation} c^{\dagger}{|n\rangle} = [n+1]^{1/2} {|n+1\rangle},\end{equation}
\begin{equation} c{|n\rangle} = [n]^{1/2} {|n-1\rangle},\end{equation}
\begin{equation} N{|n\rangle} = n{|n\rangle}.\end{equation}
We can rewrite the general commutation relations through the following  general relation % the cases of deformation through the linear second 
%order homogenous difference equation 
\cite{aba}
\be \label{e4} [N+2]=\alpha[N+1]+\beta[N],\ee
where $\alpha$ and $\beta$ depend on the \textit{$n$-bracket} definition and \textit{initial} conditions: $[0]=0$ and $[1]=1$. 
In the case of definition (\ref{e3}), we have $\alpha=q_{1}^{2}+q_{2}^2$ and $\beta=-q_{1}^2q_{2}^2$. On the other hand, 
regardless of the definition of the \textit{basic number} the relation (\ref{e4}) satisfies the Fibonacci relation. 
We can write a sequence in the form 
\begin{equation} [0] = 0, \nonumber\\\end{equation}
\begin{equation} [1] = 1, \nonumber\\\end{equation}
\begin{equation} [2] = \alpha, \nonumber\\\end{equation}
\begin{equation} \label{e5} [3] = \alpha^2+\beta, \end{equation}
\begin{equation} [4] = \alpha^3+2\alpha\beta, \nonumber\\\end{equation}
\begin{equation} [5] = \alpha^4+\beta^2+3\alpha^2\beta, \nonumber\\\end{equation}
\begin{equation} [6] = \alpha^5+4\alpha^3\beta+3\alpha\beta^2, \nonumber\\\end{equation}
\begin{equation} \nonumber{} \vdots \end{equation}
such that when $\alpha=\beta=1$ we obtain the famous Fibonacci sequence $0,1,1,2,3,5,8,...$ .

We will demonstrate from definition (\ref{e3}) that it is possible to describe the deformation models for $q$ or $p,q$. 
We can write the \textit{basic numbers} and their 
respective commutation relations by determining suitable values for $q_1$ and $q_2$: 
\begin{enumerate}
\item For $q_1=\sqrt{q}$ and $q_2=1$, 
\be [n]_q=\frac{q^n-1}{q-1},\qquad cc^\dagger-qc^\dagger c=1,\qquad \ee
that implies, $\alpha=1+q$ and $\beta=-q$.
\item For $q_1=\sqrt{q}$ and $q_2=\frac{1}{\sqrt{q}}$, we have symmetry 
\be [n]_q=\frac{q^n-q^{-n}}{q-q^{-1}},\ee 
\be cc^\dagger-qc^\dagger=q^{-n}\qquad\mbox{or}\qquad cc^\dagger-q^{-1}c^\dagger c=q^{n},\ee
then, $\alpha=q+q^{-1}$ and $\beta=-1$.
\item When $q_1=q$ and $q_2=1$, 
\be [n]_{q^2}=\frac{q^{2n}-1}{q^2-1},\qquad cc^\dagger-q^{2n}c^\dagger c=1,\ee
we have, $\alpha=1+q^{2}$ and $\beta=-q^2$.
\item When $q_1=q$ and $q_2=\frac{1}{q}$, we again have a symmetry
\be [n]_{q^2}=\frac{q^{2n}-q^{-2n}}{q^2-q^{-2}},\ee 
\be cc^\dagger-q^2c^\dagger c=q^{-2n},\qquad\mbox{or}\qquad cc^\dagger-q^{-2}c^\dagger c=q^{2n},\ee
and we get, $\alpha=q^2+q^{-2}$ and $\beta=-1$.
\item For the model $p,q$-deformed we take $q_1=\sqrt{q}$ and $q^2=\sqrt{p}$, 
\be [n]_{p,q}=\frac{q^{n}-p^{n}}{q-p},\ee 
\be cc^\dagger-qc^\dagger c=p^{n},\qquad\mbox{or}\qquad cc^\dagger-pc^\dagger c=q^{n},\ee
with $\alpha=q+p$ and $\beta=-pq$.
\end{enumerate}

All models presented above belong to the Fibonacci class of oscillators. In the literature we find various applications of these 
models. Depending on the chosen definition of the \textit{basic number}, we get different generalizations of thermostatistics.

%%%%%%%%%%%%%%%%%%%%%%%%%%%%%%%%%%%%%%%%%%%%%%%%%%%%%%%%%%%%%%%%%%%%%%%%%%%%%%%%%%%%%%%%%%%%%%%%%%%%%%%%%%%%%%%%%
\section{Hybrid Deformation Model}
\label{mdo}
\subsection{$\mu$-Deformation}
The $\mu$-oscillator definition \cite{jan1,amg} presents the same structure as the models presented in the previous section. 
Of course, the insertion of the parameter $\mu$ presents us with a definition of the \textit{basic number} structurally different 
from the previous definitions 
\be c^{\dagger}c=\Phi_{\mu}(N)=\frac{N}{1+\mu N},\qquad cc^{\dagger}=\Phi_{\mu}(N+1)=\frac{N+1}{1+\mu(N+1)}.\ee
\be \label{e6} c^{\dagger}c=\Phi_{\mu}(N)=[n]^{\mu}=\frac{n}{1+n\mu},\ee 
which in the limit $\mu\to 0$ we have $[n]^{\mu}=n$. The basic commutation relations for the $\mu$-oscillator are the 
same as for the $q$-oscillator as we can see in (\ref{e1.1}). We have the following different \textit{initial} conditions \cite{amg}
\be [0]^{\mu}=0,\qquad [1]^{\mu}=\frac{1}{1+\mu}\neq 1.\ee
We can also write the Fibonacci sequence in the same way as in (\ref{e5}), as long as
\be \lim_{\mu\to 0}~[1]^{\mu}=1.\ee

\subsection{Association of $q$-deformation with $\mu$-deformation}

Let us now associate $q$-deformation with  $\mu$-deformation.
In order to provide this we set our 
\textit{basic number} as follows
\be \label{e7}[n]_{q_1,q_2}^{\mu}=\frac{q_1^{2[n]^{\mu}}-q_2^{2[n]^{\mu}}}{q_{1}^{2}-q_{2}^{2}},\ee
\be \label{e8} \lim_{\mu\to 0}~[n]_{q_1,q_2}^{\mu}=[n]_{q_1,q_2}=\frac{q_1^{2n}-q_2^{2n}}{q_{1}^{2}-q_{2}^{2}},\qquad\mbox{and}\qquad 
\lim_{q_1=q_2\to 1}~[n]_{q_1,q_2}^{\mu}=[n]^{\mu}=\frac{n}{1+n\mu}.\ee
Of course, the definition given by (\ref{e7}) with the respective values adopted for $q_1$ and $q_2$ can recover the definitions  written for the models we 
have studied in the previous section. % but now they will be hybrid models, i.e. in the same way they will be associated with the 
%$\mu$ parameter.

The commutation relations of the hybrid model and the \textit{initial} conditions are given by
\be cc^{\dagger}-q_{1}^2c^{\dagger}c=q_{2}^{2[n]_{q_1,q_2}^{\mu}}~,\qquad\mbox{or}\qquad
cc^{\dagger}-q_{2}^2c^{\dagger}c=q_{1}^{2[n]_{q_1,q_2}^{\mu}}~,\ee
\be [0]_{q_1,q_2}^{\mu}=0,\qquad\quad [1]_{q_1,q_2}^{\mu}=\frac{q_{1}^{\frac{2}{1+\mu}}-q_{2}^{\frac{2}{1+\mu}}}{q_{1}^2-q_{2}^2}.\ee
We have 
\be \alpha=\frac{q_{1}^{\frac{4}{1+2\mu}}-q_{2}^{\frac{4}{1+2\mu}}}{q_{1}^2-q_{2}^2},\qquad\mbox{and}\qquad 
\beta=\frac{q_{1}^{\frac{6}{1+3\mu}}-q_{2}^{\frac{6}{1+3\mu}}}{q_{1}^2-q_{2}^2}-\alpha^2.\ee

Our proposal presents %as main characteristic the independence between the two models, it is easy to notice in 
%(\ref{e8}). This independence does not interfere with the insertion of deformation into a system. We have 
an association between two different types of deformation, which increases our power of application to describe a system, because  we have inserted  a new
disturbance factor (or disorder)  whose role can be played by the $\mu$-deformation in addition to the usual $q$-deformation.
% , we do not know what we can find, however, when we remove one of the 
%deformations that we leave the other still present, only when we remove the two deformations, we take the system to its initial state.

Since the deformation parameters $q_1, q_2$ or $\mu$ are independent, we further increase our manipulation power to address impurities and disorders of a physical system.  
More specifically, we can imagine, e.g., a nano film deposition system (sputtering or cathodic evaporation is used extensively in industry and in 
semiconductor research for the deposition of thin films of various materials) \cite{ches}, in which several parameters are controllable, such as
the temperature of the substrate where the material is deposited, the power of the DC source, the working 
pressure, the insertion of a gas during the process, among others. It is possible to associate the adjustable parameters 
with the deformation parameters  to calculate the thermodynamic quantities of the films, e.g.,  thermal and 
electrical conductivity. %This would not only save material time-saving material, because the films are deposited in a 
%vacuum chamber. 

%\textbf{Acho que seria interessante chamar a atenção para o fato de associarmos um modelo de osciladores de Fibonacci (e que podemos 
%descrever os modelos de deformação conhecidos na literatura) com um modelo de deformação que não obedece aos osciladores de 
%Fibonacci. E que nosso modelo ao inserir os parâmetros de deformação permite manipulá-los de forma independente: podemos reduzir para 
%apenas q; podemos usar a simetria; e ainda podemos inserir os 3 ao mesmo tempo, e ter um tipo de sequência de Fibonacci generalizada 
%e quando se retira o $\mu$ obtemos a sequência novamente.}

%\textbf{Acho que com o exemplo de uma aplicação experimental já fica suficiente. Como você sabe vender o peixe muito melhor do eu, espero 
%que estas informações ajudem.}
%%%%%%%%%%%%%%%%%%%%%%%%%%%%%%%%%%%%%%%%%%%%%%%%%%5
\section{Conclusions}
\label{con}

In this work we show that through the Fibonacci oscillators, we can write the models for one ($q$) or two ($p,q$) deformation 
parameters. And we can generically write the Fibonacci sequence, which as we know is a linear combination that encompasses 
arithmetic and geometric progressions, that is a well-known way to describe a sequence of integers. The choice of which model 
to use depends greatly on the physical system being investigated. 

%The work on $\mu$-deformation \cite{amg} that we briefly described, but we highlight the points that motivated us to propose 
We have proposed a new model of deformation,  the hybrid deformation, where we associate ($q_1,q_2$)-deformation with $\mu$-deformation. 
%We look for a way to write 
%our model, where 
This means that at the same time we insert three parameters of deformation, and that at the limit of $q_1=q_2\to 1$ we 
have only the parameter $\mu$ playing the role of deformation in the system. On the other hand, at the limit of $\mu\to 0$ we recover the usual Fibonacci oscillators. 
%Of course, that with these limits being considered at the same time, we have no deformation applied. 
%Another important feature is that, as we have seen, with the appropriate values for $q_1$ and $q_2$ we recover the $\mu$-deformation.% associated with the other 
%models in the literature.

The search for generalized statistics makes possible to solve problems where the standard statistic \cite{patt,hua} does not 
work. The results obtained with the insertion of the $\mu$-deformation enlarge our field of investigation, % because, when comparing with 
%standard results it is possible to associate as we suggest to
such as investigating factors of disorder or impurities in a thermodynamic system.

%%%%%%%%%%%%%%%%%%%%%%%%%%%%%%%%%%%%%%%%%%%%%%%%%%%%%%%%%%%%%%%%%%%%%%%%%%%%%%%
\section*{Acknowledgments}

We would like to thank CNPq, CAPES, and PNPD/PROCAD-CAPES, for partial financial support. FAB acknowledges support from CNPq (Grant no. 312104/2018-9).
%%%%%%%%%%%%%%%%%%%%%%%%%%%%%%%%%%%%%%%%%%%%%%%%%%%%%%%%%%%%%%%%%%%%%%%%%%%%%%%%%%%%%%%%%%%%%%

\end{document}